\documentclass[aps,preprint,showpacs,preprintnumbers,amsmath,amssymb]{revtex4}

\usepackage{dcolumn}% Align table columns on decimal point
\usepackage{bm}% bold math
\usepackage{amsmath,amssymb,epsfig,float}

\begin{document}

\title{Hawking radiation, Entanglement and Teleportation in background of an asymptotically
flat static black hole}

\author{Qiyuan Pan \ \ \ \ Jiliang  Jing }
\thanks{Corresponding author, Email: jljing@hunnu.edu.cn}
\affiliation{ Institute of Physics and  Department of Physics,
\\ Hunan Normal University, Changsha, \\ Hunan 410081, P. R. China
\\ and
\\ Key Laboratory of Low-dimensional Quantum Structures
\\ and Quantum
Control of Ministry of Education, \\ Hunan Normal University,
Changsha, Hunan 410081, P. R. China}

\vspace*{0.2cm}
\begin{abstract}
\vspace*{0.2cm}

The effect of the Hawking temperature on the entanglement and
teleportation for the scalar field in a most general, static and
asymptotically flat black hole with spherical symmetry has been
investigated. It is shown that the same ``initial entanglement" for
the state parameter $\alpha$ and its ``normalized partners"
$\sqrt{1-\alpha^{2}}$ will be degraded by the Hawking effect with
increasing Hawking temperature along two different trajectories
except for the maximally entangled state. In the infinite Hawking
temperature limit, corresponding to the case of the black hole
evaporating completely, the state has no longer distillable
entanglement for any $\alpha$. It is interesting to note that the
mutual information in this limit equals to just half of the
``initially mutual information". It has also been demonstrated that
the fidelity of teleportation decreases as the Hawking temperature
increases, which just indicates the degradation of entanglement.

\end{abstract}

\vspace*{1.5cm}

\maketitle

\section{introduction}

The quantum information theory in the relativistic framework has
received considerable attention due to its theoretical importance
and practical application \cite{Peres,Boschi,Pan}. Especially, more
and more efforts have been expended on the study of quantum
entanglement in a relativistic setting because people consider the
entanglement to be a major resource for quantum information tasks
such as quantum teleportation, quantum computation and so on
\cite{Bouwmeester}. With the intention of studying the entanglement
between accelerated observers, the fidelity of teleportation between
two parties in relative uniform acceleration was discussed by Alsing
\emph{et al.} \cite{Alsing-Milburn,Alsing-McMahon-Milburn}. Xian-Hui
Ge \emph{et al.} extended the gravitational field of the
teleportation to the four and higher dimensional spacetimes, and
even explicitly discussed what effects the shape of the cavity in
which particles are confined has on the teleportation in a black
hole spacetime \cite{Ge-Shen,Ge-Kim}. In order to further
investigate the observer-dependent character of the entanglement,
Fuentes-Schuller \emph{et al.} analyzed the entanglement between two
modes of a non-interacting  massless scalar field when one of the
observers describing the state is uniformly accelerated
\cite{Schuller-Mann}. And then Alsing \emph{et al.} calculated the
entanglement between two modes of a free Dirac field described by
relatively accelerated parties in a flat spacetime
\cite{Alsing-Mann}. Their results \cite{Schuller-Mann,Alsing-Mann}
also showed that the different type of field will have a
qualitatively different effect on the degradation of entanglement
produced by the Unruh effect \cite{Davies,unruh}. More recently, Ahn
\emph{et al.} extended the investigation to the entanglement of a
two-mode squeezed state in Riemannian spacetime \cite{Ahn-Kim}, Yi
Ling \emph{et al.} discussed the entanglement of electromagnetic
field in noninertial reference frames \cite{Ling}, and Adesso
\emph{et al.} investigated the distribution of entanglement between
modes of a free scalar field from the perspective of observers in
uniform acceleration \cite{Adesso}.

As a further step along this line, we will provide an analysis of
the entanglement for the scalar field in the spacetime of a most
general, static and asymptotically flat black hole with spherical
symmetry. It seems to be an interesting study to consider the
influences of the Hawking effect
\cite{Hawking-1,Hawking-2,Hawking-3} on the quantum entangled states
and show how the Hawking temperature will change the properties of
the entanglement and teleportation. Choosing a generically entangled
state as the initially entangled state for two observers in the flat
region of this black hole, we will also try to see what effects the
uncertain entangled state will have on the degradation of
entanglement in our scheme due to the presence of an arbitrary state
parameter. Our scheme proposes that the two observers, Alice and
Bob, share an initially entangled state at the same initial point in
flat Minkowski spacetime before the black hole is formed. After the
coincidence of Alice and Bob, Alice stays stationary at the
asymptotically flat region, while Bob falls in toward the mass and
then hovers outside of it. Once Bob is safely hovering outside of
the object at some constant acceleration, let it collapse to form a
black hole. By Birkhoff's theorem \cite{Birkhoff} this won't change
the metric outside of the black hole and therefore won't change
Bob's acceleration. Thus, Bob's detector registers only thermally
excited particles due to the Hawking effect \cite{unruh-1,unruh-2}.
In order to investigate the teleportation between two modes of a
scalar field as detected by the two observers, we assume that Alice
and Bob each hold an optical cavity which is small and perfect for
the teleportation in the black hole spacetime. Just as suggested by
Refs. \cite{Alsing-Milburn,Alsing-McMahon-Milburn}, we further
suppose that each cavity supports two orthogonal modes, with the
same frequency, which are each excited to a single photon Fock state
at the coincidence point for Alice and Bob. Different from the
standard teleportation protocol, our scheme assumes that Bob hovers
outside of the object before it collapses, and turns on his detector
after the formation of the black hole. Then, Bob can check to see
whether any thermal photons have been excited in his local cavity
using the non-absorbing detector.

The organization of this paper is as follows. In Sec. 2 we discuss
the vacuum structure of the background spacetime and the Hawking
effect for the scalar particles as experienced by the observer
outside the black hole. In Sec. 3 we analyze the effects of the
Hawking temperature on the entanglement between the modes for the
different state parameter. In Sec. 4 we describe the process of the
teleportation between Alice and Bob, and calculate the fidelity of
teleportation. We summarize and discuss our conclusions in the last
section.

\section{Vacuum structure and Hawking Radiation of scalar field}

It is well known that the spherically symmetric line element of a
static and asymptotically flat black hole such as Schwarzschild
black hole, Reissner-Nordstr\"{o}m black hole \cite{Chandrasekhar},
Garfinkle-Horowitz-Strominger dilaton black hole \cite{Horowitz},
Casadio-Fabbri-Mazzacurati (CFM) brane black hole \cite{Casadio} and
so on can be written in the form
\begin{eqnarray}\label{metric}
ds^2=f(r)dt^{2}-\frac{1}{h(r)}dr^{2}-R^{2}(r)(d\theta^{2}+\sin\theta^{2}d\varphi^{2}),
\end{eqnarray}
where the functions $f(r)$ and $h(r)$ vanish at the event horizon
$r=r_{+}$ of the black hole. Throughout this paper we use
$G=c=\hbar=\kappa_{B}=1$. It is obvious that the surface gravity of
the event horizon is determined by
$\kappa=\sqrt{f'(r_{+})h'(r_{+})}/2$. Defining the tortoise
coordinates $r_{*}$ as $dr_{*}=dr/\sqrt{f(r)h(r)}$, we can rewrite
the metric (\ref{metric}) as
\begin{eqnarray}\label{new metric}
ds^2=f(r)(dt^{2}-dr_{*}^{2})-R^{2}(r)(d\theta^{2}+\sin\theta^{2}d\varphi^{2}).
\end{eqnarray}

The massless scalar field $\psi$ satisfies the Klein-Gordon equation
\begin{eqnarray}
\label{K-G Equation}\frac{1}{\sqrt{-g}}\frac{{\partial}}{\partial
x^{\mu}} \left(\sqrt{-g}g^{\mu\nu}\frac{\partial\psi}{\partial
x^{\nu}}\right)=0.
\end{eqnarray}
After expressing the normal mode solution as \cite{unruh,D-R}
\begin{eqnarray}
\psi_{\omega lm}=\frac{1}{R(r)}\chi_{\omega
l}(r)Y_{lm}(\theta,\varphi)e^{-i\omega t},
\end{eqnarray}
we can easily get the radial equation
\begin{eqnarray}\label{radial equation}
\frac{d^{2}\chi_{\omega
l}}{dr_{*}^{2}}+[\omega^{2}-V(r)]\chi_{\omega l}=0,
\end{eqnarray}
with
\begin{eqnarray}
V(r)=\frac{\sqrt{f(r)h(r)}}{R(r)}\frac{d}{dr}\left[\sqrt{f(r)h(r)}\frac{d
R(r)}{dr}\right]+\frac{l(l+1)f(r)}{R^{2}(r)},
\end{eqnarray}
where $Y_{lm}(\theta,\varphi)$ is a scalar spherical harmonic on the
unit twosphere. Solving Eq. (\ref{radial equation}) near the event
horizon, we obtain the incoming wave function which is analytic
everywhere in the spacetime manifold \cite{D-R}
\begin{eqnarray}
\psi_{in,\omega lm}=e^{-i\omega v}Y_{lm}(\theta,\varphi),
\end{eqnarray}
and the outgoing wave functions for the inside and outside region of
the event horizon
\begin{eqnarray}\label{inside mode}
\psi_{out,\omega lm}(r<r_{+})=e^{i\omega u}Y_{lm}(\theta,\varphi),
\end{eqnarray}
\begin{eqnarray}\label{outside mode}
\psi_{out,\omega lm}(r>r_{+})=e^{-i\omega u}Y_{lm}(\theta,\varphi),
\end{eqnarray}
where $v=t+r_{*}$ and $u=t-r_{*}$. Eqs. (\ref{inside mode}) and
(\ref{outside mode}) are analytic inside and outside the event
horizon respectively, so they form a complete orthogonal family. In
second-quantizing the field $\Phi_{out}$ in the exterior of the
black hole we can expand it as follows \cite{unruh}
\begin{eqnarray}\label{First expand}
&&\Phi_{out}=\sum_{lm}\int d\omega[b_{in,\omega lm}\psi_{out,\omega
lm}(r<r_{+})+b^{\dag}_{in,\omega lm}\psi^{*}_{out,\omega
lm}(r<r_{+})\nonumber\\
&& \quad \quad \quad  \quad \quad \quad +b_{out,\omega
lm}\psi_{out,\omega lm}(r>r_{+})+b^{\dag}_{out,\omega
lm}\psi^{*}_{out,\omega lm}(r>r_{+})],
\end{eqnarray}
where $b_{in,\omega lm}$ and $b^{\dag}_{in,\omega lm}$ are the
annihilation and creation operators acting on the vacuum of the
interior region of the black hole, and $b_{out,\omega lm}$ and
$b^{\dag}_{out,\omega lm}$ are the annihilation and creation
operators acting on the vacuum of the exterior region respectively.
Thus, the Fock vacuum state can be defined as
\begin{eqnarray}\label{dilaton vacuum}
b_{in,\omega lm}|0\rangle_{in}=b_{out,\omega lm}|0\rangle_{out}=0.
\end{eqnarray}

Introducing the generalized light-like Kruskal coordinates
\cite{D-R,Sannan,Zhao,Birrell}
\begin{eqnarray}
&&U=-\frac{1}{\kappa}e^{-\kappa u},\quad V=\frac{1}{\kappa}e^{\kappa
v},\quad {\rm if\quad r>r_{+}};\nonumber\\
&&U=\frac{1}{\kappa}e^{-\kappa u},\quad V=\frac{1}{\kappa}e^{\kappa
v}, \quad {\rm if\quad r<r_{+}},
\end{eqnarray}
and noticing that near the event horizon
\begin{eqnarray}
r_{*}\simeq\frac{1}{\sqrt{f'(r_{+})h'(r_{+})}}\ln(r-r_{+}),
\end{eqnarray}
we can obtain a complete basis of the outgoing modes according to
the suggestion of Damour-Ruffini \cite{D-R}
\begin{eqnarray}
\psi_{I,\omega lm}=e^{\frac{\pi\omega}{2\kappa}}\psi_{out,\omega
lm}(r>r_{+})+e^{-\frac{\pi\omega}{2\kappa}}\psi^{*}_{out,\omega
lm}(r<r_{+}),
\end{eqnarray}
\begin{eqnarray}
\psi_{II,\omega
lm}=e^{-\frac{\pi\omega}{2\kappa}}\psi^{*}_{out,\omega
lm}(r>r_{+})+e^{\frac{\pi\omega}{2\kappa}}\psi_{out,\omega
lm}(r<r_{+}).
\end{eqnarray}
Thus, we can also quantize the quantum field $\Phi_{out}$ in terms
of $\psi_{I,\omega lm}$ and $\psi_{II,\omega lm}$ in the Kruskal
spacetime as
\begin{eqnarray}\label{Second expand}
\Phi_{out}=\sum_{lm}\int
d\omega&[2&\sinh(\pi\omega/\kappa)]^{-1/2}[a_{out,\omega
lm}\psi_{I,\omega lm}+a^{\dag}_{out,\omega lm}\psi^{*}_{I,\omega
lm}\nonumber \\ &+&a_{in,\omega lm}\psi_{II,\omega
lm}+a^{\dag}_{in,\omega lm}\psi^{*}_{II,\omega lm}] ,
\end{eqnarray}
where the annihilation operator $a_{out,\omega lm}$ can be used to
define the Kruskal vacuum outside the event horizon
\begin{eqnarray}\label{Kruskal vacuum}
a_{out,\omega lm}|0\rangle_{K}=0.
\end{eqnarray}

According to Eqs. (\ref{First expand}) and (\ref{Second expand}), we
obtain the Bogoliubov transformations \cite{Birrell,Barnett} for the
particle creation and annihilation operators in the black hole and
Kruskal spacetimes
\begin{eqnarray}
&&a_{out,\omega lm}=\frac{b_{out,\omega
lm}}{\sqrt{1-e^{-2\pi\omega/\kappa}}}-\frac{b^{\dag}_{in,\omega lm}}{\sqrt{e^{2\pi\omega/\kappa}-1}}, \nonumber\\
&&a^{\dag}_{out,\omega lm}=\frac{b^{\dag}_{out,\omega
lm}}{\sqrt{1-e^{-2\pi\omega/\kappa}}}-\frac{b_{in,\omega
lm}}{\sqrt{e^{2\pi\omega/\kappa}-1}}.
\end{eqnarray}
We assume that the Kruskal vacuum $|0\rangle_{K}$ is related to the
vacuum of the black hole $|0\rangle_{in}\otimes|0\rangle_{out}$ by
\begin{eqnarray}\label{two vacuum}
|0\rangle_{K}=\Upsilon(b_{in,\omega lm},b^{\dag}_{in,\omega
lm},b_{out,\omega lm},b^{\dag}_{out,\omega
lm})|0\rangle_{in}\otimes|0\rangle_{out}.
\end{eqnarray}
From $[b_{in,\omega lm},b^{\dag}_{in,\omega lm}]=[b_{out,\omega
lm},b^{\dag}_{out,\omega lm}]=1$ and Eq. (\ref{Kruskal vacuum}), we
get \cite{unruh,Ahn}
\begin{eqnarray}
\Upsilon\propto \exp({b^{\dag}_{out,\omega lm}b^{\dag}_{in,\omega
lm}}~e^{-\pi\omega/\kappa}).
\end{eqnarray}
After properly normalizing the state vector, we obtain the Kruskal
vacuum which is a maximally entangled two-mode squeezed state
\cite{Barnett,Ahn}
\begin{eqnarray}\label{Scalar-vacuum}
|0\rangle_{K}=\sqrt{1-e^{-2\pi\omega/\kappa}}
\sum_{n=0}^{\infty}e^{-n\pi\omega/\kappa}|n\rangle_{in}\otimes|n\rangle_{out},
\end{eqnarray}
and the first excited state
\begin{eqnarray}\label{Scalar-excited}
&&|1\rangle_{K}=a^{\dag}_{out,\omega lm}|0\rangle_{K}\nonumber\\
&&=(1-e^{-2\pi\omega/\kappa})\sum_{n=0}^{\infty}
\sqrt{n+1}~e^{-n\pi\omega/\kappa}|n\rangle_{in}\otimes|n+1\rangle_{out},
\end{eqnarray}
where $\{|n\rangle_{in}\}$ and $\{|n\rangle_{out}\}$ are the
orthonormal bases for the inside and outside region of the event
horizon respectively. For the observer outside the black hole, he
needs to trace over the modes in the interior region since he has no
access to the information in this causally disconnected region.
Therefore, when an outside observer travels through the Kruskal
particle vacuum $|0\rangle_{K}$ of mode $\omega$ his detector
registers a number of particles given by
\begin{eqnarray}\label{Hawking}
_{K}\langle0|b^{\dag}_{out,\omega lm}b_{out,\omega
lm}|0\rangle_{K}=\frac{1}{e^{2\pi\omega/\kappa}-1}=\frac{1}{e^{\omega/T}-1},
\end{eqnarray}
where we have defined the Hawking temperature as
\cite{Kerner-Mann,Jiang-Wu-Cai}
\begin{eqnarray}\label{Hawking temperature}
T=\frac{\kappa}{2\pi}=\frac{\sqrt{f'(r_{+})h'(r_{+})}}{4\pi}.
\end{eqnarray}
Eq. (\ref{Hawking}) is well known as the Hawking effect
\cite{Hawking-1,Hawking-2,Hawking-3}, which shows that the observer
in the exterior of the black hole detects a thermal Bose-Einstein
distribution of particles as he traverses the Kruskal vacuum.

\section{Quantum Entanglement in background of a black hole}

Now we assume that Alice has a detector which only detects mode
$|n\rangle_{A}$ and Bob has a detector sensitive only to mode
$|n\rangle_{B}$, and they share a generically entangled state at the
same initial point in flat Minkowski spacetime before the black hole
is formed
\begin{eqnarray}\label{initial}
|\Psi\rangle=\alpha|0\rangle_{A}|0\rangle_{B}
+\sqrt{1-\alpha^{2}}|1\rangle_{A}|1\rangle_{B},
\end{eqnarray}
where $\alpha$ is some real number which satisfies
$|\alpha|\in(0,1)$, $\alpha$ and $\sqrt{1-\alpha^{2}}$ are the
so-called ``normalized partners". After the coincidence of Alice and
Bob, Alice remains at the asymptotically flat region but Bob freely
falls in toward the mass with his detector and then hovers outside
of it before it collapses to form a black hole. Obviously, there
will be some thermal effects due to changes in Bob's acceleration,
but there will be no Hawking radiation. Note that such effects can
be negligible, or at least will disperse after some time. Then, once
Bob is safely hovering outside of the object at some constant
acceleration, let it collapse to form a black hole. By Birkhoff's
theorem \cite{Birkhoff} this won't change the metric outside of the
black hole and therefore won't change Bob's acceleration. Thus,
Bob's detector registers only thermally excited particles due to the
Hawking effect \cite{unruh-1,unruh-2}. The states corresponding to
mode $|n\rangle_{B}$ must be specified in the coordinates of the
black hole in order to describe what Bob sees in this curved
spacetime. Thus, using Eqs. (\ref{Scalar-vacuum}) and
(\ref{Scalar-excited}), we can rewrite Eq. (\ref{initial}) in terms
of Minkowski modes for Alice and black hole modes for Bob. Since Bob
is causally disconnected from the interior region of the black hole,
we will take the trace over the states in this region and obtain the
mixed density matrix between Alice and Bob in exterior region
\begin{eqnarray}\label{Bos-density}
&&\rho_{AB}=(1-e^{-\omega/T})\sum_{n=0}^{\infty}\rho_{n}~e^{-n\omega/T},
\nonumber\\&& \rho_{n}=\alpha^{2}|0n\rangle\langle0n|
+(n+1)(1-\alpha^{2})(1-e^{-\omega/T})|1(n+1)\rangle\langle1(n+1)|
\nonumber\\&&\qquad+\alpha\sqrt{(n+1)(1-\alpha^{2})
(1-e^{-\omega/T})}|0n\rangle\langle1(n+1)|\nonumber\\&&\qquad+
\alpha\sqrt{(n+1)(1-\alpha^{2})(1-e^{-\omega/T})}
|1(n+1)\rangle\langle0n|,
\end{eqnarray}
where $|nm\rangle=|n\rangle_{A}|m\rangle_{B,out}$.

It should be noted that we do not trace over the states located
inside the event horizon for Alice, even though she now is also
causally disconnected from the interior region of the black hole. As
a matter of fact, the causal structure of spacetime keeps every
observer exterior from the black hole disconnected from its
interior. Why do we not trace over the degree of freedom of Alice in
the region inaccessible to her? We now present the reasons as
follows. On the one hand, we can justify what we are doing by
theory. For a Schwarzschild black hole, the mass of which is assumed
to be of the order of a solar mass ($M_{bh}\sim M_{\odot}$), the
magnitude of acceleration near this black hole that Bob needs is
about $10^{13}m/s^{2}$, which is much larger than that of Alice
needs (almost equals to zero) in the asymptotical region. Thus, we
argue that Alice's acceleration effects can be neglected, whereas
Bob's can't. On the other hand, we can think about it from
experiment. Though we do not observe the black hole directly,
impressive progress in optical, radio and $X$-ray astronomy greatly
bolster the evidence for supermassive black holes in the centers of
galaxies \cite{Frolov-Novikov}. Thus, the Earth can be argued to be
an asymptotical region far from black holes, as far as we know. The
standard quantum field theory works fine for the earthbound
experiments, so we have at least some circumstantial empirical
evidence that tracing over the black hole interior can be neglected
in asymptotical regions.

\begin{figure}[ht]
\includegraphics[scale=0.8]{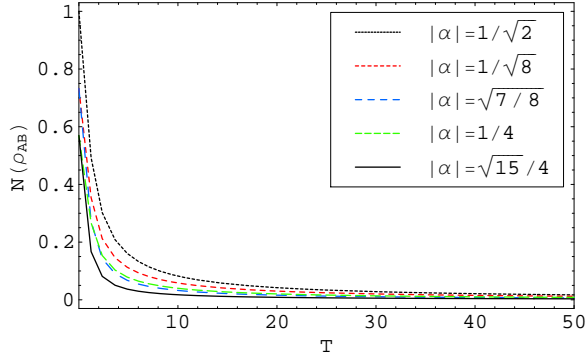}
\caption{\label{NegBos}The logarithmic negativity as a function of
the Hawking temperature $T$ with the fixed $\omega$ for different
$\alpha$.}
\end{figure}

It is clear that the partial transpose criterion provides a
sufficient condition for the existence of entanglement in this case
\cite{peres}: if at least one eigenvalue of the partial transpose of
the density matrix is negative, the density matrix is entangled; but
a state with positive partial transpose can still be entangled. It
is well-known bound or nondistillable entanglement
\cite{Vidal,Plenio}. Interchanging Alice's qubits
$(|mn\rangle\langle pq| \to |pn\rangle\langle mq|)$, we get the
matrix representation of the partial transpose in the ($n$,~$n+1$)
block {\small
\begin{eqnarray}
(\rho_{AB}^{T_{A}})_{n,n+1}&=&e^{-n\omega/T}(1-e^{-\omega/T})
\nonumber \\ && \times \left(
\begin{array}{ll}
 n(1-\alpha^{2})(e^{\omega/T}-1)
 & \alpha\sqrt{(n+1)(1-\alpha^{2})(1-e^{-\omega/T})} \\ \\
 \alpha\sqrt{(n+1)(1-\alpha^{2})(1-e^{-\omega/T})}
 & ~~~~~~~~\alpha^{2}e^{-\omega/T}\\
\end{array}\right),
\end{eqnarray}}
and its eigenvalues
\begin{eqnarray}
\lambda _{\pm }^{n}=\frac{e^{-n\omega/T}(1-e^{-\omega/T})}{2}
\left[\zeta_{n}\pm\sqrt{\zeta_{n}^{2}
+4\alpha^{2}(1-\alpha^{2})(1-e^{-\omega/T})}~\right],
\end{eqnarray}
where
$\zeta_{n}=\alpha^{2}e^{-\omega/T}+n(1-\alpha^{2})(e^{\omega/T}-1)$.
Obviously the eigenvalue $\lambda_{-}^{n}$ is always negative for
finite value of the Hawking temperature. Hence, this mixed state is
always entangled for any finite value of $T$. It should be noted
that in the limit $T\rightarrow\infty$, the negative eigenvalue will
go to zero. In order to discuss this further, we will use the
logarithmic negativity which serves as an upper bound on the
entanglement of distillation \cite{Vidal,Plenio}. This entanglement
monotone is defined as $N(\rho_{AB})=\log
_{2}||\rho_{AB}^{T_{A}}||$, where $||\rho_{AB}^{T_{A}}||$ is the
trace norm of the partial transpose $\rho_{AB}^{T_{A}}$. Thus, we
obtain the logarithmic negativity for this case
\begin{eqnarray}
&&N(\rho_{AB})=\log_{2}\left[\alpha^{2}(1-e^{-\omega/T})
+\sum_{n=0}^{\infty}e^{-n\omega/T}(1-e^{-\omega/T})
\sqrt{\zeta_{n}^{2} +4\alpha^{2}
(1-\alpha^{2})(1-e^{-\omega/T})}\right].\nonumber \\
\end{eqnarray}
The trajectories of the logarithmic negativity $N(\rho_{AB})$ versus
$T$ for different $\alpha$ in Fig. \ref{NegBos} just show how the
Hawking temperature $T$ would change the properties of the
entanglement.

For the Hawking temperature of zero, corresponding to the case of a
supermasive or an almost extreme black hole,
$N(\rho_{AB})=\log_{2}(1+2|\alpha|\sqrt{1-\alpha^{2}})$. In the
range $0<|\alpha|\leq1/\sqrt{2}$ the larger $\alpha$, the stronger
the ``initial entanglement"; but in the range
$1/\sqrt{2}\leq|\alpha|<1$ the larger $\alpha$, the weaker the
``initial entanglement". For finite Hawking temperature, the
monotonous decrease of $N(\rho_{AB})$ with increasing $T$ for five
different $\alpha$ means that the ``initial entanglement" is lost to
the thermal fields generated by the Hawking effect. This result
agrees well with the Hawking's original argument
\cite{Hawking-1,Hawking-2,Hawking-3}, which says that smaller black
holes are at a higher temperature and so radiate more violently than
massive black holes. Fig. \ref{NegBos} also shows that when the
``initial entanglement" is stronger, we lose it more rapidly. But it
is surprisingly found that the same ``initial entanglement" for
$\alpha$ and its ``normalized partner" $\sqrt{1-\alpha^{2}}$ will be
degraded along two different curves except for the maximally
entangled state, i.e., $|\alpha|=1/\sqrt{2}$. This phenomenon, due
to the coupling of $\alpha$ and the exponential functions related to
$T$, just shows the inequivalence of the quantization for a scalar
field in the black hole and Kruskal spacetimes. The logarithmic
negativity is exactly zero for any $\alpha$ in the limit
$T\rightarrow\infty$, which indicates that the state has no longer
distillable entanglement for the arbitrary values of $\alpha$ when
the black hole evaporates completely.

\begin{figure}[ht]
\includegraphics[scale=0.8]{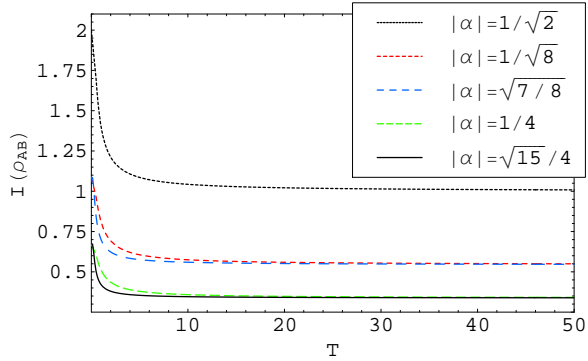}
\caption{\label{MuInBos}The mutual information as a function of the
Hawking temperature $T$ with the fixed $\omega$ for different
$\alpha$.}
\end{figure}

In order to estimate the total amount of correlations between Alice
and Bob, we will analyze the mutual information which is defined as
\cite{RAM}
\begin{eqnarray}
I(\rho_{AB})=S(\rho_{A})+S(\rho_{B})-S(\rho_{AB}),
\end{eqnarray}
where $S(\rho )=-\text{Tr}(\rho \log_{2}\rho)$ is the entropy of the
density matrix $\rho$. From Eq. (\ref{Bos-density}), we can give the
entropy of this joint state
\begin{eqnarray}
&&S(\rho _{AB})=-\sum_{n=0}^{\infty }
e^{-n\omega/T}(1-e^{-\omega/T})\left[\alpha^{2}+(n+1)(1-\alpha^{2})(1-e^{-\omega/T})\right]
\nonumber\\
&&\qquad\qquad~\times
\log_{2}e^{-n\omega/T}(1-e^{-\omega/T})\left[\alpha^{2}+(n+1)(1-\alpha^{2})(1-e^{-\omega/T})\right].
\end{eqnarray}
Tracing over Alice's states for the density matrix $\rho_{AB}$, we
get Bob's density matrix in exterior region of the event horizon
\begin{eqnarray}
&&\rho_{B}=(1-e^{-\omega/T})\sum_{n=0}^{\infty}e^{-n\omega/T}\left[\alpha^{2}|n\rangle\langle
n|+(n+1)(1-\alpha^{2})(1-e^{-\omega/T})|n+1\rangle\langle
n+1|\right],
\end{eqnarray}
and its entropy
\begin{eqnarray}
&&S(\rho_{B})=-\sum_{n=0}^{\infty
}e^{-n\omega/T}(1-e^{-\omega/T})\left[\alpha^{2}+n(1-\alpha^{2})(e^{\omega/T}-1)\right]
\nonumber\\
&&\qquad\qquad\times
\log_{2}e^{-n\omega/T}(1-e^{-\omega/T})\left[\alpha^{2}+n(1-\alpha^{2})(e^{\omega/T}-1)\right].
\end{eqnarray}
We can also obtain Alice's density matrix by tracing over Bob's
states
\begin{eqnarray}\label{Alice-density}
&&\rho_{A}=\alpha^{2}|0\rangle\langle0|
+(1-\alpha^{2})|1\rangle\langle1|,
\end{eqnarray}
whose entropy can be expressed as
\begin{eqnarray}\label{Alice-entropy}
&&S(\rho_{A})=-[\alpha^{2}\log_{2}\alpha^{2}
+(1-\alpha^{2})\log_{2}(1-\alpha^{2})].
\end{eqnarray}
Thus, we draw the behaviors of the mutual information $I(\rho_{AB})$
as a function of the Hawking temperature $T$ for different values of
the state parameter $\alpha$ in Fig. \ref{MuInBos}.

Fig. \ref{MuInBos} shows that for the Hawking temperature of zero,
the ``initially mutual information" equals to
\begin{eqnarray}\label{Boc-MuIn-Initial}
I_{i}(\rho _{AB})=-2[\alpha^{2}\log_{2}\alpha^{2}
+(1-\alpha^{2})\log_{2}(1-\alpha^{2})].
\end{eqnarray}
In the range $0<|\alpha|\leq1/\sqrt{2}$ the larger $\alpha$, the
stronger $I_{i}(\rho _{AB})$; but in the range
$1/\sqrt{2}\leq|\alpha|<1$ the larger $\alpha$, the weaker
$I_{i}(\rho _{AB})$. As the Hawking temperature increases, the
mutual information becomes smaller. It is interesting to note that
except for the maximally entangled state, the same ``initially
mutual information"  for $\alpha$ and $\sqrt{1-\alpha^{2}}$ will be
degraded along two different trajectories. However, in the infinite
Hawking temperature limit $T\rightarrow\infty$, i.e., the black hole
evaporates completely, the mutual information converges to the same
value again
\begin{eqnarray}\label{Boc-MuIn-Final}
I_{f}(\rho _{AB})=-[\alpha^{2}\log_{2}\alpha^{2}
+(1-\alpha^{2})\log_{2}(1-\alpha^{2})],
\end{eqnarray}
which equals to just half of $I_{i}(\rho _{AB})$. Thus, we conclude
that
\begin{eqnarray}
&&I_{f}(\rho _{AB})=\frac{1}{2}I_{i}(\rho _{AB}),
\end{eqnarray}
which is independent of the state parameter $\alpha$. Obviously if
$I_{i}(\rho _{AB})$ is higher, it is degraded to a higher degree in
this limit. Since the distillable entanglement in the infinite
Hawking temperature limit is zero, we are safe to say that the total
correlations consist of classical correlations plus bound
entanglement in this limit.

\section{Quantum Teleportation in background of a black hole}

In this section we will concentrate on a particular quantum
information task: quantum teleportation. We assume that Alice and
Bob each hold an optical cavity, at rest in their local frame. Each
cavity supports two orthogonal modes (labeled by $A_{i}$ and $B_{i}$
with $i=1,~2$), with the same frequency, which are each excited to a
single photon Fock state at the coincidence point for Alice and Bob.
We ignore the polarization of these modes and model the photons by
the massless modes of a scalar field as suggested by Refs.
\cite{Alsing-Milburn,Alsing-McMahon-Milburn}. Considering the
textbook teleportation protocol \cite{Nielsen}, we let Alice and Bob
share a maximally entangled state, i.e., an entangled Bell state in
flat Minkowski spacetime
\begin{eqnarray}
|\Psi\rangle=\frac{1}{\sqrt{2}}\left(|\mathbf{0}\rangle_{A}|\mathbf{0}\rangle_{B}
+|\mathbf{1}\rangle_{A}|\mathbf{1}\rangle_{B}\right),
\end{eqnarray}
where the logical states $|\mathbf{0}\rangle_{A}$ and
$|\mathbf{1}\rangle_{A}$ are defined in terms of the physical Fock
states for Alice's cavity by the dual-rail basis states
\cite{Alsing-Milburn,Alsing-McMahon-Milburn}
\begin{eqnarray}
|\mathbf{0}\rangle_{A}=|1\rangle_{A_{1}}|0\rangle_{A_{2}},~~
|\mathbf{1}\rangle_{A}=|0\rangle_{A_{1}}|1\rangle_{A_{2}},
\end{eqnarray}
with similar expressions for Bob's cavity. It should be noted that
$|1\rangle_{A_{1}}$ and $|1\rangle_{A_{2}}$ are single photon
excitations of the Minkowski vacuum states in Alice's cavity. Our
construction implicitly assumes that we have chosen a modal
decomposition of the Minkowski vacuum based on intra-cavity and
extra-cavity modes, which is a legitimate alternative to the usual
way of quantizing the vacuum in terms of plane wave modes
\cite{Dalton-Knight-1,Dalton-Knight-2}. Once the cavities are loaded
with a photon, we also assume each cavity is perfect and cannot emit
the photon.

Recalling the usual teleportation protocol with the unknown state
\cite{Nielsen}
\begin{eqnarray}\label{teleportation}
|\varphi\rangle=a|\mathbf{0}\rangle+b|\mathbf{1}\rangle,
\end{eqnarray}
we assume that Alice has an additional cavity which contains this
single qubit (\ref{teleportation}) with dual-rail encoding by a
photon excitation of a two-mode Minkowski vacuum state. This will
allow Alice to make a joint measurement on the two orthogonal modes
of each cavity. For the usual teleportation protocol between two
Minkowski observers Alice and Bob, after Alice's measurement, Bob's
state will be projected according to the measurement outcome. We can
give the final state received by Bob
\begin{eqnarray}
|\varphi_{ij}\rangle=x_{ij}|\mathbf{0}\rangle+y_{ij}|\mathbf{1}\rangle,
\end{eqnarray}
with four possible conditional state amplitudes
$(x_{00},y_{00})=(a,b)$, $(x_{01},y_{01})=(b,a)$,
$(x_{10},y_{10})=(a,-b)$ and $(x_{11},y_{11})=(-b,a)$. Once
receiving the classical information of the result of Alice's
measurement, Bob can apply a unitary transformation to verify the
protocol. Obviously the fidelity of the teleported state is unity in
this idealized situation.

Alice now wishes to perform the same teleportation protocol with the
noninertial observer Bob. We assume that prior to their coincidence,
Alice and Bob ensure that all photons are removed from their
cavities. When Alice and Bob instantaneously share a maximally
entangled state at the asymptotically flat region, we suppose that
the two cavities overlap and simultaneously a four photon source
excites a two photon state in each cavity. Then Alice remains there
but Bob falls in toward the mass and then hovers outside of it. Once
Bob is safely hovering outside of the object at some constant
acceleration, let it collapse to form a black hole. Then, Bob turns
on his detector after the formation of the black hole. Bob can check
to see whether any thermal photons have been excited in his local
cavity using the non-absorbing detector. It should be noted that the
common frequency of both Alice's and Bob's cavity is just the
frequency $\omega$ of Eqs. (\ref{Scalar-vacuum}) and
(\ref{Scalar-excited}) \cite{Alsing-Milburn,Alsing-McMahon-Milburn}.
For Bob, the observer locates near the event horizon of a black
hole, he needs to trace over the modes in the interior region since
he is causally disconnected from this region. Thus, when Alice sends
the result of her measurement to Bob, Bob's state can be projected
into
\begin{eqnarray}\label{Bob teleport}
&&\rho_{ij}=\sum_{k=0}^{\infty}\sum_{l=0}^{\infty}~_{in}\langle
k,l|\varphi_{ij}\rangle\langle\varphi_{ij}|k,l\rangle_{in}\nonumber\\
&&~~~~=(1-e^{-\omega/T})^{3}\sum^{\infty}_{n=0}\sum^{n}_{m=0}\{e^{-(n-1)\omega/T}[(n-m)|x_{ij}|^{2}+m
|y_{ij}|^{2}]|m,n-m \rangle_{out}\langle m,n-m|
\nonumber\\&&\qquad~~+[
\sqrt{(m+1)(n-m+1)}~x_{ij}y^{*}_{ij}e^{-n\omega/T}|m,n-m+1\rangle_{out}\langle
m+1,n-m|+{\rm H.c.}]\},\nonumber \\
\end{eqnarray}
where $|m,n-m \rangle_{out}=|m\rangle_{B_{1}}\otimes|n-m
\rangle_{B_{2}}$ is a state of $n$ total excitations in the exterior
region product state, with $0\leq m\leq n$ excitations in the
leftmost mode. Eq. (\ref{Bob teleport}) can be rewritten as
\begin{eqnarray}
&&\rho_{ij}=\sum^{\infty}_{n=0}p_{n}\rho_{ij,n},\quad {\rm
with}\quad p_{0}=0,\quad
p_{n}=(1-e^{-\omega/T})^{3}e^{-(n-1)\omega/T}\quad{\rm for}\quad
n\geq1.
\end{eqnarray}
Since what we concern about is to which extent
$|\varphi_{ij}\rangle$ might deviate from unitarity, so upon
receiving the result $(i,j)$ of Alice's measurement, Bob can apply
the rotation operators (a unitary transformation in his local frame)
restricted to the one-excitation sector of his state spanned by
$\{|\mathbf{0}\rangle_{out},|\mathbf{1}\rangle_{out}\}=\{|0,1\rangle_{out},|1,0\rangle_{out}\}$
to turn this portion of his density matrix into the exterior region
analogue of the state in Eq. (\ref{teleportation})
\cite{Alsing-Milburn,Alsing-McMahon-Milburn,Ge-Shen,Ge-Kim}
\begin{eqnarray}
|\varphi\rangle_{out}=a|\mathbf{0}\rangle_{out}+b|\mathbf{1}\rangle_{out}.
\end{eqnarray}
Thus, we can obtain the fidelity of Bob's final state with
$|\varphi\rangle_{out}$
\begin{eqnarray}
F\equiv{}~_{out}\langle\varphi|\rho_{ij}|\varphi_{ij}\rangle_{out}
=(1-e^{-\omega/T})^{3}.
\end{eqnarray}

\begin{figure}[ht]
\includegraphics[scale=0.8]{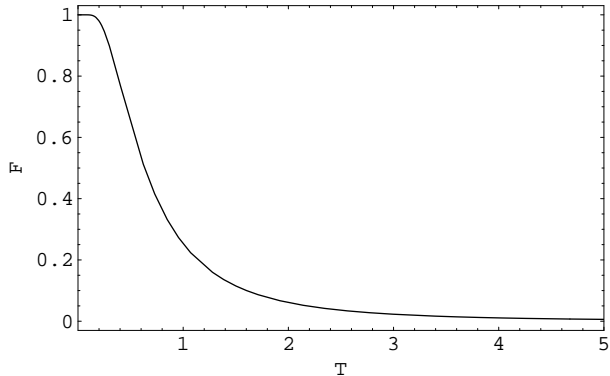}
\caption{\label{FBosw}The fidelity of teleportation as a function of
the Hawking temperature $T$ with the fixed $\omega$ for a maximally
entangled state.}
\end{figure}

From Fig. \ref{FBosw}, we can see that the fidelity of teleportation
depends on the Hawking temperature $T$. It has been found that the
fidelity decreases as the Hawking temperature increases, which just
indicates the entanglement degradation obtained in previous section
because the state fidelity in conventional teleportation protocol is
related to the entanglement.

\section{summary}

We have analytically discussed the effect of the Hawking temperature
on the entanglement between two modes of a scalar field as detected
by Alice who stays stationary at an asymptotically flat region and
Bob who locates near the event horizon in the background of a most
general, static and asymptotically flat black hole with spherical
symmetry. It is shown that the entanglement is degraded by the
Hawking effect with increasing Hawking temperature. It is found that
the stronger the ``initial entanglement" which corresponds to the
Hawking temperature of zero, i.e., the case of a supermasive or an
almost extreme black hole, the faster it loses. It is found that the
same ``initial entanglement" for the state parameter $\alpha$ and
its ``normalized partners" $\sqrt{1-\alpha^{2}}$ will be degraded
along two different trajectories as the Hawking temperature
increases except for the maximally entangled state
$\alpha=1/\sqrt{2}$, which just shows the inequivalence of the
quantization for a scalar field in the black hole and Kruskal
spacetimes. In the infinite Hawking temperature limit
$T\rightarrow\infty$, corresponding to the case of the black hole
evaporating completely, the state has no longer distillable
entanglement for the arbitrary values of $\alpha$. Further analysis
shows that the mutual information is degraded to a nonvanishing
minimum value which is dependent of $\alpha$ with increasing Hawking
temperature. However, it is interesting to note that the mutual
information in the infinite Hawking temperature limit equals to just
half of the ``initially mutual information", which is independent of
$\alpha$. We have also investigated the scheme of teleportation in
this black hole spacetime. It has been demonstrated that the
fidelity of teleportation decreases as the Hawking temperature
increases, which just indicates the entanglement degradation because
the state fidelity in conventional teleportation protocol is related
to the entanglement.

\begin{acknowledgments}
The authors would like to thank the anonymous referee for his/her
insightful and constructive criticisms and suggestions, which
allowed us to improve the manuscript significantly. This work was
supported by the National Natural Science Foundation of China under
Grant No. 10675045; the FANEDD under Grant No. 200317; and the Hunan
Provincial Natural Science Foundation of China under Grant No.
08JJ3010.

\end{acknowledgments}

\end{document}